

\def\v{{\bf v}}
\def\eps{\varepsilon}
\def\p{{\bf p}}
\def\pp{{\bf p}^\prime}
\def\r{{\bf r}}
\def\q{{\bf q}}
\def\om{\omega }
\def\bra{\big\langle}
\def\ket{\big\rangle}
\def\argwig{(\r,\p,t)}
\def\parnep{{\partial n^0 \over \partial \eps _p}}
\def\hbo{\hbar \om}


\magnification=\magstep1
\baselineskip 15pt plus 2pt

\font\headl=cmfib8 scaled \magstep3
\parindent=20pt
\topskip=2truecm

\nopagenumbers
\centerline{{\headl Quantum effects in the stochastic behaviour of}}
\medskip
\centerline{{\headl nuclear matter at finite excitations\footnote{$^*$}
{$\rm Dedicated\; to\; the\; 60^{\rm th}\; birthday\; of\; Professor\; K.\;
Dietrich$}
}}
\bigskip\bigskip
\bigskip
\centerline{by}
\bigskip\bigskip\bigskip
\centerline{\bf Dieter Kiderlen and Helmut Hofmann}
\centerline{Physik-Department, TU M\"unchen}
\centerline{D-85747 Garching}
\vskip 2cm

\par{\narrower\noindent
We examine the dynamics of statistical fluctuations in nuclear matter.
Linear response functions for the average phase space density are
derived within Landau theory. Properties of the stochastic forces are
deduced from the quantal fluctuation dissipation theorem, with a
suitable generalization to unstable modes. Sizable quantum effects are
found both inside and outside the spinodal regime.
\par}
\vfill
\eject
\topskip=10pt
\pageno=1 \footline={\rm \hfil \folio \hfil}

\ref{bertsiem}
\edef\nbertsiem{\the\refnumber}
\ref{heis}
\edef\nheis{\the\refnumber}
\ref{lalif}
\edef\nlalif{\the\refnumber}
\ref{hofkitse}
\edef\nhofkitse{\the\refnumber}
\ref{hofthoming}
\edef\nhofthoming{\the\refnumber}
\ref{hofing}
\edef\nhofing{\the\refnumber}
\break
{\it \num Introduction:}
The dynamics of fluctuations has become an area of wide interest in
nuclear physics. In this paper we concentrate on examples
which are related to spinodal instabilities. In [\nbertsiem] the latter
have been suggested as a possible origin for the fragmentation of a hot
nuclear system formed in a heavy ion collision at intermediate energies.
Ideally, one would like to have a realistic model to study explicitly
the time evolution of nuclear matter. The simplest possibility is
offered by the average one body density. Indeed, in a series
of papers Landau theory has been exploited to obtain this information,
see [\nheis].

An obvious next step is to include fluctuations. For this one might
think to proceed as described in [\nlalif] for hydrodynamics of a stable
system: Take the basic equations for average dynamics and add
fluctuating forces to construct a coupled set of Langevin equations.
The Langevin force can be fixed by its relation to the fluctuations in
thermal equilibrium, which allows the inclusion of quantum effects.

Rather than hydrodynamics we prefer to choose Landau theory as formulated
in [\nheis]. Moreover, we do not use the Langevin equation itself, but
address directly the time evolution of the second moments (cumulants)
around the average field. Such an approach has been formulated in
[\nhofkitse]. Different to that, the present version is not restricted to
stable modes. This allows to treat the dynamics of quantal statistical
fluctuations even in the regime of spinodal instabilities.

One may wonder whether such quantum effects make sense at all as the
Landau-Vlasov equation itself does not contain $\hbar$ explicitly. We
are, however, encouraged by the results of [\nhofthoming] (see also
[\nhofing]). There the decay of a meta-stable state across a
one-dimensional barrier has been studied using the same method as
suggested here. In this way one has been able to deduce results known
before from exact treatments of model systems with functional integrals.
\ref{pinoz}
\edef\npinoz{\the\refnumber}
\ref{baympeth}
\edef\nbaympeth{\the\refnumber}
\ref{abrikha}
\edef\nabrikha{\the\refnumber}
\medskip
{\it \num Average dynamics from Landau theory:}
Our main goal is the equation of motion for the fluctuations.
A basic ingredient for their construction are response functions deduced
from average dynamics. These can be obtained
from a linearized equation for the phase space density. Within Landau
theory it can be written as [\nlalif, \npinoz, \nbaympeth, \nheis]:
\equorder{eqofmotrel}
$$ {\partial\over\partial t}\delta n\argwig + \v_p\nabla_r\delta n\argwig -
    {\partial n^0\over\partial\eps_p} \v_p\nabla_r\delta \eps\argwig
     + {n\argwig -  n_\p^{\rm l.e.}(\eps_\p)\over\tau} =0
    \eqna $$
\edef\neqofmotrel{\the\equnumbera}
The $\delta n\argwig$ is the deviation of the actual density
$n\argwig= \bra {\hat n}(\r ,\p) \ket_t$  from its value $n^0(\eps^0_p)$
in equilibrium for which we take the grand canonical distribution with
a chemical potential $\mu_0$ and temperature $T_0$. The $\delta \eps\argwig$
measures the deviation of the actual quasi-particle energy $\eps\argwig $ from
the equilibrium value $ \eps^0_p$. It is influenced both by the external
field $\delta U\argwig$ and the actual density:
\equorder{deltaeps}
$$\delta \eps\argwig = {1\over {\cal V}}\sum_{\pp}
   f_{\p \pp} \delta n(\r ,\pp ,t) + \delta U\argwig \eqna $$
\edef\ndeltaeps{\the\equnumbera}
Here $f_{\p \pp}$ are the Landau forces. In contrast to conventional
procedures we allow $\delta U$ to depend on $\r$ and $\p$. The last
term on the left of (\neqofmotrel) represents the collision term in
relaxation time approximation.  The $n^{\rm l.e.}$ is the density
distribution of the local equilibrium.  To linear order it is given by
\equorder{loceqgleq}
$$ n^{\rm l.e.}(\eps_p)=n^0(\eps^{0}_p) +
    {\partial n^0\over\partial\eps_p}\left(\delta\eps_p-\p\delta {\bf u}-
    \delta\mu + {\eps^{0}_p-\mu_0\over T_0}\delta T\right) \eqna $$
\edef\nloceqgleq{\the\equnumbera}
One would like to have eq.(\neqofmotrel) satisfy the basic conservation laws.
This is possible by an appropriate choice of $\delta {\bf u}$ (representing
the local flow velocity), $\delta\mu$ and  $\delta T$  as functionals of
$\delta n_p$ and $\delta \eps_p$ [\nabrikha, \nbaympeth, \nheis].

To simplify the actual computation we make a kind of low temperature
approximation: The factor $-\parnep$ appearing in (\neqofmotrel) is
approximated by $\delta (\eps^0_p - \mu^0)$ everywhere but not for the
calculation of the density of states $N(T)$ at the Fermi surface (see below).
Essentially, this implies neglect of heat transport [\nheis].

\ref{forster}
\edef\nforster{\the\refnumber}
\medskip
{\it \num Linear response functions:}
We parametrize the solution $\delta n$ of (\neqofmotrel) in terms of
response functions $\chi$, namely those which determine how the density
$n$ changes with the external field $\delta U$.
It is convenient to use Fourier transforms from $\r$ to $\q$ and
$t$ to $\om$ [\nhofkitse]. The $\chi$ is then defined by
\equorder{deltanres}
$$\delta n(\q_1,\p_1,\om)=  -{1\over {\cal V}} \sum_{\q_2,\p_2} \;
     \chi_{\p_1,\p_2}(\q_1,\q_2,\om)  \; \delta U(\q_2,\p_2,\om)  \eqna $$
\edef\ndeltanres{\the\equnumbera}
It must be proportional to $\delta (\q_1+\q_2)$, if the unperturbed
equilibrium state is assumed  to be homogeneous in space [\nforster,
\npinoz].  Thus, it suffices to study functions
$\chi_{\p_1,\p_2}(\q,\om)$ which depend on $\q={1\over 2}(\q_1-\q_2)$
only. Their equation of motion can be obtained  from eq.(\neqofmotrel)
by taking the functional derivative with respect to $\delta
U(\q,\p_2,\om)$. One obtains:
\equorder{eqofmotchi}
$$(\om-\v_{p_1}\q)\chi_{\p_1,\p_2}(\q,\om)+
   {\partial n^0\over\partial\eps_{p_1}} \v_{p_1}\q
   \left({1\over {\cal V}}\sum_{\pp}
    f_{\p_1 \pp} \chi_{\pp,\p_2}-h^3\delta(\p_1-\p_2)\right) -
    {i\over\tau}{\delta(n_{p_1}-n^{\rm l.e.}_{p_1})\over\delta U(\p_2)}
    =0 \eqna $$
\edef\neqofmotchi{\the\equnumbera}
To evaluate the last term on the right hand side one may proceed as
follows. First one  uses (\nloceqgleq) to write $n_p-n^{\rm l.e.}_p$ as
functional of $\delta n_p$ and $\delta\eps_p$ (c.f.[\nheis]). Then one
exploits (\ndeltaeps) together with the definition of the response
function.

\ref{vankam}
\edef\nvankam{\the\refnumber}
A closer look at eq.(\neqofmotchi) allows to find terms proportional to
$\delta(\p_1-\p_2)$ additional to the one already shown.
This enables one to write the general solution of (\neqofmotchi) in the
following form:
\equorder{resvarcov}
$$\chi_{\p_1,\p_2}(\q,\om)=(2\pi \hbar)^3\delta (\p_1-\p_2)
  \chi^{\rm var}_{\p_1}(\q,\om) + \chi^{\rm cov}_{\p_1,\p_2}(\q,\om) \eqna $$
\edef\nresvarcov{\the\equnumbera}
where all terms which are regular in $\p_1,\p_2$ are put together in
$\chi^{\rm cov}_{\p_1,\p_2}$. In this way one recognizes a structure of the
response functions which has proven convenient for the case of equilibrium
fluctuations (see e.g. [\nvankam]), namely the splitting into a variance and
a covariance part. Indeed, this feature is of particular interest for our
purpose, as finally we will calculate the equilibrium fluctuations from our
response functions.

{}From eq.(\neqofmotchi) one may deduce equations of motion for $\chi^{\rm
cov}$
and $\chi^{\rm var}$. The one for $\chi_\p^{\rm var}(\q,\om)$ turns out
particularly simple. It is worth while to show it explicitly:
\equorder{eomresvar}
$$\left(-i\om + i\v_p \q +{1\over \tau}\right) \chi^{\rm var}_{\p}(\q,\om) =
    - \parnep \left(i\v_p \q + {1\over \tau}\right)   \eqna $$
\edef\neomresvar{\the\equnumbera}
We return to it below.
\goodbreak
\medskip
{\it \num Dynamics of fluctuations:}
For $\delta n \equiv \bra\delta{\hat n}\ket_t$, the fluctuations are defined
as:
$\sigma_{\p_1,\p_2}(\q_1,\q_2,t)= {1\over2}\bra
\left[\delta{\hat n}(\q_1,\p_1), \delta{\hat n}(\q_2,\p_2))\right]_+
\ket_t - \bra \delta {\hat n}(\q_1 ,\p_1) \ket_t
          \bra \delta {\hat n}(\q_2,\p_2) \ket_t$.
According to [\nhofkitse] the equations of motion for $\sigma (t)$
take the form
\footnote{$^1$}{The procedure utilized here is similar to the one
suggested in Vol. X of [\nlalif] to evaluate macroscopic (two-time)
correlation functions.}
:
\equorder{eomfluct}
$$\left({\partial\over\partial t} +
   ({\cal D}_1+{\cal D}_2) \ast \right)\sigma_{\p_1,\p_2}(\q_1,\q_2,t) =
   2 d(\q_1,\p_1;\q_2,\p_2)  \eqna $$
\edef\neomfluct{\the\equnumbera}
if the operator ${\cal D}$ is defined by writing eq.(\neqofmotrel) as:
$\left(\partial / \partial t +{\cal D}_1 \ast \right)
\delta n(\q_1,\p_1,t)) = 0$.
The star is introduced to stress that ${\cal D}$ does not simply stand for
differential operators, but that it represents summations (or integrations)
over momenta as well, as can be seen from eq.(\ndeltaeps).
The left hand side of (\neomfluct) is entirely determined by (\neqofmotrel).
The quantity on the right hand side
simulates the presence of stochastic forces, which for the present,
linearized situation may be assumed to be represented just by the
constant $2 d$ (being independent of the $\sigma$'s).

\medskip
{\it \num Diffusion coefficients from fluctuation dissipation theorem:}
{}From the structure of eq.(\neomfluct) we expect, due to the
collision term, that $\sigma_{\p_1,\p_2}(\q_1,\q_2,t)$ will show a
relaxational behaviour. For the moment let us look at stable situations.
As $t \to \infty$ the $\sigma_{\p_1,\p_2}(\q_1,\q_2,t)$ relaxes
to its equilibrium value $\sigma^{\rm eq}_{\p_1,\p_2}(\q)$
(which like $\chi$ depends on $\q$ only). This allows to
{\it actually calculate the diffusion coefficients from those}:
$2d(\q_1,\p_1,\q_2,\p_2)=
({\cal D}_1+{\cal D}_2)*\sigma^{\rm eq}_{\p_1,\p_2}(\q_1,\q_2)$,
for $\delta U = 0$, of course.
So all we need to do is to calculate the
$\sigma^{\rm eq}_{\p_1,\p_2}(\q)$, for instance from the fluctuation
dissipation theorem (FDT). As we will see this is possible also in the
regime of instabilities where the notion of an equilibrium fluctuation
looses its  physical meaning.  Let us proceed in two steps and explain
first our application of FDT to stable modes, and then discuss the
generalization to unstable ones.

Restricting to the dependence on $\q$, the consequences of the FDT on
the equilibrium fluctuations may be stated as
(see Vol IX of [\nlalif] and [\npinoz, \nforster]):
\equorder{fdt}
$$  \sigma^{\rm eq}_{\p_1,\p_2}(\q) = \hbar \int_C \;
   {d\omega\over 2\pi}\;\;\coth\left({\hbo\over 2T}\right)
   \chi_{\p_1,\p_2}^{\prime\prime}(\q,\omega) \eqna $$
\edef\nfdt{\the\equnumbera}
with the contour $C$ representing integration along the real axis from
$-\infty$ to $\infty$, for stable modes. The $\chi^{\prime\prime}$ is
the dissipative part of the response function,
$2 i\chi^{\prime\prime}_{1,2}(\q,\om) =
\chi_{1,2}(\q,\om) - \chi_{2,1}(-\q,-\om)$.

An unstable system is characterized by the appearance of a pole of
$\chi_{1,2}(\q,\om)$ in the upper half of the $\om$-plane.
Certainly, we would like to be able to treat this case as an analytic
continuation of the stable one, simply because one moves from one
condition to the other one by changing smoothly some physical
conditions. Mathematically this condition requires to deform the contour
C such that all poles of the integrand remain on the same side of C, as
they have been for the corresponding stable case.
This rule can be obeyed if C is chosen i) to lie above all poles of
$\chi_{1,2}(\q,\om)$ and below the ones of $\chi_{2,1}(-\q,-\om)$, and
ii) to cross the imaginary axis in between the two the Matsubara
frequencies which lie closest to the real axis, namely $\om_{\pm 1}=\pm
2\pi T/\hbar$. This latter condition puts a lower limit on the range
of $T$: our treatment is bound to fail for
$ T < T_0 \equiv \hbar\om^{\rm max}_q / 2\pi$, where
$\om_q^{\rm max}$ refers to the most unstable mode.

That such a procedure makes sense can be learned from the case of one
collective variable as discussed in [\nhofthoming, \nhofing], where the
integral in (\nfdt) can be carried out analytically in terms of the
logarithmic derivative $\psi$ of the $\Gamma$-function. It can easily
be checked that for the unstable mode the result is identical to the
one of the prescription just described, where (\nfdt) is evaluated
along the contour $C$.
\medskip
{\it \num Application to the variance part}:
In (\nresvarcov) we have noted the particular dependence of the
response functions on the two momenta $\p_1, \p_2$. Because of the very
construction through the FDT (\nfdt) a similar behaviour will hold true
for both the equilibrium fluctuations and the diffusion coefficients.
The form of eq.(\neomfluct) implies that
$\sigma_{\p_1,\p_2}(\q_1,\q_2,t)$ itself can then be split into a variance
and a covariance part, just because the inhomogeneity $2d$ has
this structure. Similar to the case of the response function, the equation
for $\sigma_\p^{\rm var}(\q_1,\q_2,t)$ is easily derived by identifying in
(\neomfluct) all terms proportional to $\delta(\p_1-\p_2)$. One obtains:
\equorder{eomfluctvar}
$$ \left({\partial\over\partial t}+ i\v_p (\q_1 + \q_2) + {2\over
    \tau}\right)   \sigma^{\rm var}_{\p}(\q_1,\q_2,t)=
    (2\pi \hbar)^3\delta(\q_1+\q_2) \; 2d_{\rm var}(\q,\p) \eqna $$
\edef\neomfluctvar{\the\equnumbera}
Again, $d_{\rm var}$ is to be calculated from the left side in the
stationary limit and via the generalized FDT (\nfdt), which explains
the special structure of the right hand side.

For a constant relaxation time $\tau $ the integral in (\nfdt) diverges
in the quantum regime as can be seen solving (\neomresvar). The
reason for this is the Markov approximation for the collision term in
(\neqofmotrel) which becomes invalid for short times, say on a
time scale $1/\om_D$.  To improve the approximation one has to take
into account memory effects in the collision process. The simplest
possible way is to put $\tau (\om )=\tau (1-i\om/\om_D)$, a procedure
which usually is called Drude regularization. It can be considered as
the first step in the hierarchy of approximations writing $1/\tau$ as a
continued fraction and truncating the latter by putting $\om_D$ a
constant rather than a function of $\om$. In our practical
calculations we have chosen the value $\om_D = 100 \; {\rm MeV}$ and we
have convinced ourselves that in the range of $\om_D= 30 \cdots 200
\;{\rm MeV}$ our results of the diffusion coefficient change less than
about $20\% $.

The regularization is not necessary in the high temperature limit which
is defined by replacing in (\nfdt)
$\coth\left({\hbo\over 2T}\right)$ by ${2T\over \hbo}$. It can be
noticed, that in this way the $\hbar$ drops out, which indicates that
this is the limit of classical physics. Using the
properties of the Fermi function one gets from
eqs.(\neomfluctvar), (\nfdt) and (\neomresvar):
\equorder{diffvar}
$$ d^{\rm high T}_{\rm var}(\q,\p)=
     {1\over\tau_0}n^0(\eps^0_p)(1-n^0(\eps^0_p))  \eqna $$
\edef\ndiffvar{\the\equnumbera}
\ref{lhpan}
\edef\nlhpan{\the\refnumber}
\ref{danielo}
\edef\ndanielo{\the\refnumber}
\ref{koehl}
\edef\nkoehl{\the\refnumber}
{\it \num Choice of the microscopic forces:}
For the Landau force a form as in [\nheis] is used, i.e.
$f_{\p \p^\prime} = f_0+\beta q^2 +f_1 \hat\p \cdot \hat\p^\prime$.
The parameters $f_0, f_1$ are related to the compressibility $\kappa$
and the effective mass, respectively (see [\nbaympeth]). The latter
are determined by fitting the Skyrme-type interaction of [\nlhpan] to
the Friedman-Pandharipande free energy for symmetric nuclear matter.
In the relation between $f_0$ and $\kappa$ the q-p density of states
$N(T)$ enters. It has been taken to be $T$-dependent to determine $f_0$
reasonably well in the spinodal region. The term $\sim q^2$ takes
into account the effect of the inhomogeneity in the presence of a
density wave [\nheis]. The value $\beta=50 \; {\rm MeV\; fm^5}$ is
adjusted to reproduce the spinodal region as given in [\nheis].

For the relaxation time $\tau$ we have used two microscopic results, as
in [\ndanielo] and [\nkoehl]. The one of [\ndanielo] is based on an
estimate of the shear viscosity $\eta $ (see the discussion in [\nheis]).
In [\nkoehl] relaxation times $\tau _l$ are given for deformations of the
Fermi surface of various multipolarities $l$. We took their analytical
formulas for $\tau_2$ as function of the space density $\rho$ and
temperature. The restriction to $l=2$ is done for reasons of simplicity,
first of all. Secondly, according to [\nbaympeth] it is this $\tau_2$
for which the association to the shear viscosity can be made.
\ref{cochora}
\edef\ncochora{\the\refnumber}

\medskip
{\it \num Numerical results:}
In the following we exhibit the effects discussed above by numerical
computations. As an example let us consider the diffusion coefficient
$d_{2,2}$, obtained by projecting on quadrupole distortions of the Fermi
surface according to the general expression (see [\nhofkitse])
\equorder{defdlk}
$$ d_{lk} (\q)=(2l+1)(2k+1){1\over {\cal V}^2}\sum_{\p_1,\p_2}
   P_l(\hat p_1\hat q) P_k(\hat p_2\hat q) d(\q,\p_1,\p_2) \eqna $$
\edef\ndefdlk{\the\equnumbera}

The results are plotted in Figs.1 and 2 for different temperatures
as function of $\rho /\rho_0$ (with $\rho_0$ being the saturation
density). The value taken for the wave number $q$ is the same as used
in [\ncochora] and is compatible with the range of wave numbers
discussed in [\nheis].
When the density decreases one crosses the isothermal spinodal line.
In Fig.1 the points at which that happens are marked by a vertical bar.
It can be seen that our results cross these points in a
{\it completely smooth} way, which is a consequence of our choice of $C$
in the integral (\nfdt).

In Fig.1 we compare the results for $d_{2,2}$ obtained from the full,
quantal version of (\nfdt) (fully drawn lines) with the corresponding
results for the high $T$-limit, which for the present approximation  
can be written as: $d_{lk}^{\rm high T}=\delta_{l,k} T N(T) (2l+1)
(1-\delta_{l,0}-\delta_{l,1})/\tau $. The interesting feature is that quantum
effects are present up to $T \simeq 10 {\rm MeV}$. In Fig.2 we present
computations done for our two choices of the relaxation time.  In each
case we plot the results for both $d_{2,2}$ and the variance part
$d_{2,2}^{\rm var}$.  Three features can be seen: i) The two models for
the relaxation times lead to quite similar results.  ii) The {\it
covariance} part plays a role only for small temperatures.  iii) It is
this part which is responsible for the structure of the curves seen at
lower temperatures.

\ref{kidthese}
\edef\nkidthese{\the\refnumber}
\ref{brenigtwo}
\edef\nbrenigtwo{\the\refnumber}
\ref{randrem}
\edef\nrandrem{\the\refnumber}
\ref{burgchorand}
\edef\nburgchomrand{\the\refnumber}
\ref{benidhernrem}
\edef\nbenidhernrem{\the\refnumber}
\medskip
{\it \num Discussion:}
We have presented a method which allows to include the effects of
quantal statistical fluctuations on top of a description for the
average density. The procedure is based on a linearization around
an instantaneous equilibrium which permits to take advantage of the
FDT to determine the stochastic forces. For realistic applications
it is necessary though to generalize this FDT to include unstable
modes. One of the benefits we see in this technique is the fact that
the treatment of average dynamics may well rely on a phenomenological
approach, like the one of Landau theory taken here.

Our procedure can be considered a generalization of the case of one
collective variable (plus its conjugate momentum) presented in
[\nhofthoming, \nhofing] (and in earlier publications cited therein)
to the case of infinitely many variables on which
the one body distribution function $n(\r,\p,t)$ depends in principle.
The discussion in
[\nhofthoming] offers a direct proof that the technique applied there
is capable of treating quantum effects of global motion, in as much as,
for a model case, relations have been established to the analytically
solvable example of "dissipative tunneling". In that case a condition
on temperature was found which is stronger than the one we have
encountered above when constructing the contour $C$, namely
$T>T_c=2T_0$. Furthermore, for $T<T_c$ some of the diffusion coefficients
become negative in that case,
but this problem has so far not been encountered in
our present work. In the present case $T_0$ turns out smaller than
about $1.43 \; {\rm MeV}$ (attained at $\rho=0.045\, {\rm fm^{-3}}$).
As this value is close to the one (1.5 MeV) chosen for the lower
curves in Fig. 1,2, we would not like to put too much emphasis on the
peculiar feature found there.
Its nature will have to be clarified in the future.

One striking result of our present computation is the rather large
value of $T$ up to which quantum effects prevail. According to Fig.2,
at $T=12 \; {\rm MeV}$ the quantal and the classical diffusion
coefficient still differ by a factor of about $1.5$. This difference
comes entirely from the influence of the factor
$\coth\left({\hbo\over 2T}\right)$, as compared to the ${2T\over \hbo}$
in the classical limit. This is possible only because the contribution
of $\chi_{\p_1,\p_2}^{\prime\prime}(\q,\omega)$ extends to somewhat
larger values of $\om$. More details will be provided in a forthcoming
paper [\nkidthese].

Of course, our present computations do not yet allow to draw definite
conclusions on the general $T$-dependence of our transport coefficients.
For instance, with respect to the application of Landau theory itself
we have made approximations discussed at the end of sect.2 which first
have to be improved.

In this context we mention a particular subtlety of the
FDT, in its application to the computation of fluctuations. In (\nfdt)
the integrand represents the correlation function associated with
$\chi_{\p_1,\p_2}^{\prime\prime}(\q,\omega)$. The former may have a
contribution of the type $\Delta \sigma \delta (\om)$. The constant
$\Delta \sigma $ can be fixed by the precise physical conditions
under which the fluctuations are to be calculated. To clarify this
point it suffices to look at the high temperature limit, for which,
in our case, $\sigma^{\rm eq}_{\p_1,\p_2}(\q)$ is given by $T$ times
the {\it static response} $\chi_{\p_1,\p_2}(\q,\om=0)$. Instead, for
situations where the system is coupled to a real heat bath, one ought
to evaluate $\sigma^{\rm eq}$ from the
{\it isothermal susceptibility} $\chi^T_{\p_1,\p_2}(\q)$, which can
be done by choosing $\Delta \sigma_{\p_1,\p_2} (\q) =
T (\chi^T_{\p_1,\p_2}(\q) - \chi_{\p_1,\p_2}(\q,\om=0))$
(see e.g.[\nbrenigtwo]).

However, we feel that this term cannot be of great importance here.
First of all, for a real, finite nuclear system a heat bath in
{\it true sense} does not exist, although one may well use the
concept of temperature to parametrize excitations. Of course, for an
isolated system the entropy $S$ would be more appropriate, in principle,
and one might thus think of changing (\nfdt) by adding a term as discussed
above, but with $\chi^T$ replaced by the adiabatic susceptibility
$\chi^S$. However, then an argument involving ergodicity comes into
play: Assuming the latter to be given for a nuclear system, one knows
$\chi^S- \chi(\om=0)$ to vanish in such a case [\nbrenigtwo].
Secondly, for our model case of nuclear matter, it is only consistent
with our general approximations to put $\chi^T-\chi(\om=0)=0$.
For an ergodic system this relation becomes correct at $T=0$
[\nbrenigtwo] and, as described in sect.2, we used only the low
temperature version of Landau theory.

It is reassuring to have several examples for which our calculation of
the $ \sigma $'s can be checked independently by comparison with other
derivations. For instance, for the free gas we get the leading term of
the fluctuations in the phase space density (see sect.8 of [\nhofkitse]),
and for an interacting system it is possible to recover ([\nkidthese])
the correct expression for the density fluctuations,  both calculated
in the long wavelength limit $q \to 0$.
Next let us look at the conservation of the total number of particles,
a condition which may be put as a sum rule
(see e.g. (20.16) of Vol.X of [\nlalif]). In our context the latter
can be written as:  $\sigma_{0,k}(q=0)=0$ for all $k$. As will be
shown in [\nkidthese], at least for the case of forces with a finite
number of non vanishing Landau parameters, this equation is compatible
with our general results for the fluctuations.

Moreover, in the high temperature limit we are able to establish a
relation of our treatment to the one of [\nrandrem]. This is possible
by comparing  the variance part of our diffusion coefficient
(\ndiffvar) to the one derived in [\nrandrem]
(called ${1\over 2}\alpha^2[f]$ in [\nburgchomrand])
and applied in [\ncochora] to a situation close to equilibrium. In
this model the relaxation time $\tau$ may be expressed by the
transition rates for the gain and loss term, $W^+$ and $W^-$,
respectively ([\nbenidhernrem]). Then
${1\over 2}\alpha^2[n^0]$ is easily seen to have the same form as
given by the right hand side of (\ndiffvar). Of course, in order to
make the identity complete one must reinterpret the density of the
quasi-particles of Landau theory as the particle density in the sense
of the Boltzmann equation.
A comparison of the covariance parts is more lengthy, but
the result is of the same kind as for the variance parts: the
high temperature limit of $2d_{\rm cov}$ equals the quantity
$\alpha_{\rm cov}[n^0]$ of [\nburgchomrand], if only the calculation is
performed consistently with the approximation used for the collision term.
Moreover, the sum rule (17) of [\nburgchomrand] is found to be fulfilled
by our results even in the general case.
More details will be provided in [\nkidthese].

\bigskip
{\it Acknowledgments}
One of us (D.K.) wants to thank the Niels Bohr Institute for the kind
hospitality extended to him during his stay in Copenhagen. We would
like to thank H. Heiselberg and C. Pethick for valuable suggestions
and discussions and M.M. Sharma and F.J.W. Hahne for carefully reading the
manuscript.

\goodbreak
\vskip 0.5truecm
\parindent=20pt

\leftline{{\it REFERENCES}}
\medskip

\item{1)} G F. Bertsch and P.J. Siemens, Phys. Lett. 126B (1983) 9
\item{2)} H. Heiselberg, C.J. Pethick and D.G. Ravenhall,
          Ann.Phys. (N.Y.) 223 (1993) 37;
          {\it ibid} Phys. Rev. Lett. 61 (1988) 818
\item{3)} L.D. Landau und E.M. Lifschitz, Course of Theoretical Physics,
          Vol. IX, "Statistical Physics, part 2" and
          Vol. X, "Physical Kinetics", E.M. Lifschitz und I.P. Pitajewski,
          Pergamon (1980) Oxford
\item{4)} H. Hofmann, D. Kiderlen and I. Tsekhmistrenko,
          Z.Phys.A 341 (1992) 181
\item{5)} H. Hofmann, M. Thoma and G.-L. Ingold, to be published
\item{6)} H. Hofmann and G.-L. Ingold, Phys. Lett. 264B (1991) 253
\item{7)} D. Pines and P. Nozi\`eres, "The Theory of Quantum Liquids",
          W.A. Benjamin, Inc., New York, 1966
\item{8)} G. Baym and C.J. Pethick, "Landau Fermi Liquid Theory",
          J. Wiley \& Sons (1991), New York
\item{9)} A.A. Abrikosov and I.M. Khalatnikov, Rep. Progr. Phys. 22
          (1959) 329
\item{10)} D. Forster, "Hydrodynamic Fluctuations, Broken Symmetry and
          Correlation Functions", W.A. Benjamin, 1975, London
\item{11)} N.G. van Kampen, "Stochastic Processes in Physics and Chemistry",
           North Holland, New York, 1981
\item{12)} V. R. Pandharipande and D.G. Ravenhall, "Hot Nuclear Matter",
           Les Houches Winter School on "Nuclear Matter and Heavy Ion
           Collisions", Feb. 7-16, 1989
\item{13)} P. Danielewicz, Phys. Letts. B146 (1984) 168
\item{14)} M.M. Abu-Samreh and H.S. K\"ohler, Nucl. Phys. A552 (1993) 101
\item{15)} M. Colonna, Ph. Chomaz and J. Randrup, preprint (1993) GANIL P 93 01
\item{16)} D. Kiderlen, Doctoral Thesis (TUM), to be completed 1994
\item{17)} W. Brenig, "Statistical Theory of Heat, Nonequlibrium Phenomena",
           Springer, 1989, Berlin
\item{18)} J. Randrup and B. Remaud, Nucl. Phys. A514 (1990) 339
\item{19)} G.F. Burgio, P. Chomaz, J. Randrup, Nucl. Phys. A529 (1991) 157
\item{20)} B. Benhassine, M. Farine, E.S. Hern\'andez, D. Idier, B. Remaud
           and F. S\'ebille, preprint Universt\'e Nantes, IPN 93-06,
           to be published in Nucl. Phys.

\goodbreak
\vskip 0.5truecm
\parindent=0pt
\leftline{{\it FIGURE CAPTIONS}}
\medskip
Fig.1: Quantal diffusion coefficient $d_{2,2}$ (fully drawn lines) and
its high  temperature limit (dashed lines) as function of density (see text).

Fig.2: Quantal diffusion coefficient $d_{2,2}$ (full and dashed lines)
and its variance part (dashed-dotted and dotted lines)
for relaxation times taken from [\ndanielo] (full and
dashed-dotted lines) and from [\nkoehl] (dashed and dotted lines).
\end